\documentclass[aps,pre,floats,twocolumn,showpacs,superscriptaddress]{revtex4}

\usepackage{graphicx,epsfig}% Include figure files
\usepackage{times}
\usepackage{graphics,dcolumn,bm,fleqn,float}
\usepackage{amssymb,amsmath,multirow,rotate,color}
\begin{document}

\title{Residential segregation and cultural dissemination: An Axelrod-Schelling model.}

\author{C. Gracia-L\'azaro}

\affiliation{Departamento de F\'{\i}sica de la Materia Condensada,
University of Zaragoza, Zaragoza E-50009, Spain}

\author{L. Fern\'andez-Lafuerza}

\affiliation{IFISC, Instituto de F´\'{\i}sica Interdisciplinar y
Sistemas Complejos (CSIC-UIB), E-07122 Palma de Mallorca, Spain}

\author{L. M. Flor\'{\i}a}

\email{mario.floria@gmail.com}

\affiliation{Institute for Biocomputation and Physics of Complex
Systems (BIFI), University of Zaragoza, Zaragoza 50009, Spain}

\affiliation{Departamento de F\'{\i}sica de la Materia Condensada,
University of Zaragoza, Zaragoza E-50009, Spain}

\author{Y. Moreno}

\email{yamir.moreno@gmail.com}

\affiliation{Institute for Biocomputation and Physics of Complex
Systems (BIFI), University of Zaragoza, Zaragoza 50009, Spain}

\affiliation{Departamento de F\'{\i}sica Te\'orica. University of
Zaragoza, Zaragoza E-50009, Spain}

\date{\today}

\begin{abstract}

In the Axelrod's model of cultural dissemination, we consider
mobility of cultural agents through the introduction of a density
of empty sites and the possibility that agents in a dissimilar
neighborhood can move to them if their mean cultural similarity
with the neighborhood is below some threshold. While for low
values of the density of empty sites the mobility enhances the
convergence to a global culture, for high enough values of it the
dynamics can lead to the coexistence of disconnected domains of
different cultures. In this regime, the increase of initial
cultural diversity paradoxically increases the convergence to a
dominant culture. Further increase of diversity leads to
fragmentation of the dominant culture into domains, forever
changing in shape and number, as an effect of the never ending
eroding activity of cultural minorities.

\end{abstract}


\maketitle

%\section{Introduction}

The use of agent-based models (ABM) \cite{Axel1} in the study of
social phenomena provides useful insights about the fundamental
causal mechanisms at work in social systems. The large-scale
(macroscopic) effects of simple forms of (microscopic) social
interaction are very often surprising and generally hard to
anticipate, as vividly demonstrated by one of the earliest
examples of ABM, the Schelling \cite{Sch1,Sch2} model of urban
segregation, that shows how residential segregation can emerge
from individual choices, even if people have fairly tolerant
preferences regarding the share of like persons in a residential
neighborhood.

To gain insights on the question of why cultural differences
between individuals and groups persist despite tendencies to
become more alike as a consequence of social interactions, Axelrod
\cite{Axel2} proposed an ABM for the dissemination of culture,
that has subsequently played a prominent role in the investigation
of cultural dynamics. Questions concerning the establishment,
spread and sustainability of cultures, as well as on the "pros and
cons" of cultural globalization versus the preservation and
coexistence of cultural diversity, are of central importance both
from a fundamental and practical point of view in today's world.

The Axelrod model implements the idea that social influence is
"homophilic", {\em i.e.} {\em the likelihood that a cultural
feature will spread from an individual to another depends on how
many other features they may have already in common} \cite{Axel2}.
The resulting dynamics converges to a global monocultural
macroscopic state when initial cultural diversity is below a
critical value, while above it homophilic social influence is
unable to inforce cultural homogeneity, and multicultural patterns
persist asymptotically. This change of behavior has been
characterized \cite{castellano1,castellano2,vilone,vazquez1} as a
non-equilibrium phase transition. Subsequent studies have analyzed
the effects on this transition of different lattice or network
structures \cite{Klemm1,Klemm1a}, the presence of different types of noise
("cultural drift") \cite{Klemm2,Klemm2a}, as well as the consideration of
external fields (influential media) \cite{Gonzalez1} and global or
local non-uniform couplings \cite{Gonzalez2}. Up to now, no investigation of the effects of agent mobility on cultural transmission has been carried out, with the exception of \cite{Axtell}, where individuals move following the gradient of a "sugar" landscape (that they consume) and interact culturally with agents in their neighborhood, i.e., mobility is not culturally driven.

In this paper we incorporate into the Axelrod cultural dynamics
the possibility that agents living in a culturally dissimilar
environment can move to other available places, much in the spirit
of the Schelling model of segregation. This requires the
introduction of a density of empty sites $h$ in the discrete space
(lattice) where agents live. As anticipated by \cite{Axtell} the
expectations are that the agents mobility should enhance the
convergence to cultural globalization, in the extent that it acts
as a sort of global coupling between agents. It turns out that
these expectations are clearly confirmed when the density $h$ of
empty sites is low enough so that the set of occupied sites
percolates the lattice: The transition value depends linearly with
the number of agents, so that in an infinite system (thermodynamical limit) only
global cultural states are possible. However, for large enough
values of $h$, new phenomena appear associated to the mixed
Axelrod-Schelling social dynamics, including a new multicultural
fragmented phase at very low values of the initial cultural
diversity, a (seemingly first order) transition to cultural
globalization triggered by mobility, and the fragmentation of the
dominant culture into separated domains that change continuously
as the result of erosive processes caused by the mobility of
cultural minorities.

%\section{The model}
%\label{II}

In the Axelrod model of cultural dissemination, a culture is
modelled as a vector of $F$ integer variables $\{\sigma_f\}$
($f=1,...,F$), called cultural {\em features}, that can assume $q$
values, $\sigma_f= 0,1,...q-1$, the possible {\em traits} allowed
per feature. At each elementary dynamical step, the culture
$\{\sigma_f(i)\}$ of an individual $i$ randomly chosen is allowed
to change (social influence) by imitation of an uncommon feature's
trait of a randomly chosen neighbor $j$, with a probability
proportional to the cultural overlap $\omega_{ij}$ between both
agents, defined as the proportion of shared cultural features,
\begin{equation}
\omega_{ij} =
\frac{1}{F}\sum_{f=1}^{F}\delta_{\sigma_f(i),\sigma_f(j)}, 
\label{overlap}
\end{equation}
where $\delta_{x,y}$ stands for the Kronecker's delta which is 1 if $x=y$ and 0 otherwise. Note that in the Axelrod dynamics the mean cultural overlap
$\bar{\omega}_i$ of an agent $i$ with its $k_i$ neighbors, defined
as
\begin{equation}
\bar{\omega}_{i} = \frac{1}{k_i}\sum_{j=1}^{k_i}\omega_{ij} \;\; ,
\label{meanoverlap}
\end{equation}
not always increases after an interaction takes place with a
neighboring agent: indeed, it will decrease if the feature whose
trait has been changed was previously shared with at least two
other neighbors.

To incorporate the mobility of cultural agents into the Axelrod
model, two new parameters are introduced, say the density of empty
sites $h$, and a threshold $T$ ($0\leq T \leq 1$), that can be
called {\em intolerance}. After each elementary step of the
Axelrod dynamics, we perform the following action: If imitation
has not occurred and $\omega_{ij}\neq 1$, we compute the mean
overlap (\ref{meanoverlap}) and if $\bar{\omega}_i < T$, then the
agent $i$ moves to an empty site that is randomly chosen. Finally,
in the event that the agent $i$ randomly chosen is isolated (only
empty sites in its neighborhood), then it moves directly to an
empty site.

We define the mobility $m_i$ of an agent $i$ as the probability
that it moves in one elementary dynamical step (provided it has
been chosen):
\begin{equation}
m_i = (1-\bar{\omega}_{i})\; \Theta(T-\bar{\omega}_{i}) \;\; ,
\label{mobility}
\end{equation}
where $\Theta(x)$ is the Heaviside step function, that takes the
value 1 if $x>0$, and 0 if $x\leq 0$. For an isolated agent, that
moves with certainty, one may convene that its mean cultural
overlap is zero, so that expression (\ref{mobility}) applies as
well. The average mobility $m$ of a configuration is the average
of the mobility of the agents:

\begin{equation}
m = \frac{1}{N} \sum_{i=1}^N m_i \;\; , \label{meanmobility}
\end{equation}
where $N$ is the total number of cultural agents. We will consider
below two-dimensional square lattices of linear size $L$, so that
$N=(1-h)L^2$, periodic boundary conditions, and von Neumann
neighborhoods, so that the number $k_i$ of neighbors of an agent
$i$ is $0 \leq k_i \leq 4$. We fix the number of cultural features
to $F=10$, and vary the parameters $q$, $h$ and $T$, as well as
the linear size $L$ of the lattice.

%\section{Results and Discussion}

For the initial conditions for the cultural dynamics, $N$ cultural
agents are randomly distributed in the $L \times L$ sites of the
square lattice, and randomly assigned a culture. The simulation is
stopped when the number $n_a$ of active links ({\em i.e.}, links
such that $0 < \omega_{ij} < 1$) vanishes. The results shown below
are obtained by averaging over a large number (typically $5\cdot10^2\;
-\; 10^4$) different initial conditions.

\begin{figure}
\begin{center}
\epsfig{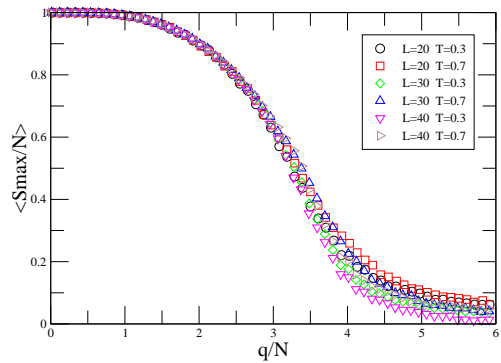}
\end{center}
\caption{Order parameter $\langle S_{max} \rangle / N$ versus
scaled initial cultural diversity $q/N$ for a very small density of
empty sites $h=0.05$ and different values of the intolerance $T=0.3,
\; 0.7$, and of the lattice linear size $L=20, \; 30, \; 40$, as
indicated in the inset.}
\label{fig1}
\end{figure}

The usual order parameter for the Axelrod model is $\langle
S_{\text {max}}\rangle /N$, where $\langle S_{\text {max}}\rangle$
is the average number of agents of the dominant (most abundant)
culture. Large values (close to unity) of the order parameter are
the signature of cultural globalization. In Fig.\ \ref{fig1}, we plot the
order parameter versus the initial cultural diversity scaled to
the population size, $q/N$, for a small value of the density of
empty sites $h=0.05$, and different values of the intolerance $T$
and of the linear size $L$. We observe the collapse in a single
curve of the graphs corresponding to different lattice sizes and,
moreover, that the results are rather insensitive to the
intolerance values.

%\end{center}

%\begin{center}
\begin{figure}
\begin{center}
\epsfig{file=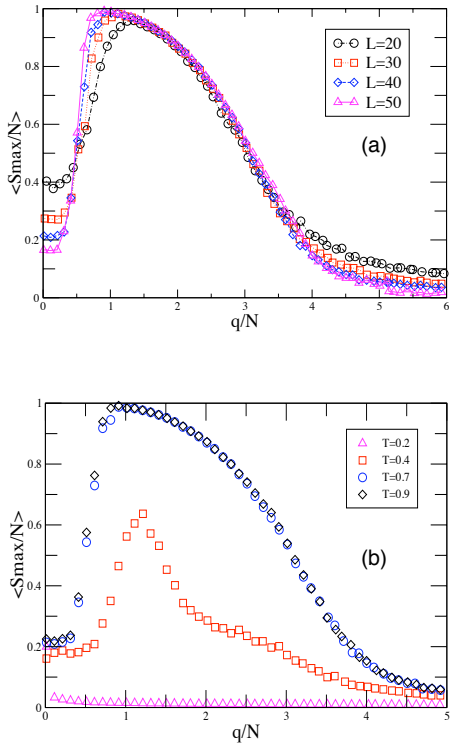, width=7cm, angle=0}
\end{center}
\caption{(color online) Order parameter $\langle S_{max} \rangle / N$ versus
scaled initial cultural diversity $q/N$ for an intermediate value of
the density of empty sites $h=0.5$. Panel (a) corresponds to a high value of the intolerance
$T=0.7$, and different lattice linear sizes $L=20, \; 30, \; 40, \;
50$, while in panel (b) $L=40$, and
different values of the intolerance $T=0.2, \; 0.4, \; 0.7, \; 0.9$ are used. See the text for further details.}
\label{fig2}
\end{figure}
%\end{center}

For a fixed value of the initial cultural diversity $q$, the
larger the size $N$ of the population is, the more likely an agent
can share a cultural feature with someone else in the population.
Hence, as mobility allows contacts with virtually anybody, the
increase of the population size enhances the tendency towards
cultural globalization, and the monocultural (ordered) phase
extends up to higher values of the parameter $q$. The critical
value $q_c$ of the transition between consensus and a disordered
multicultural phase diverges with the system size $q_c \sim N$, so
that in the thermodynamical limit only global cultural states are
possible for a small density $h$ of empty sites.

We will focuss hereafter on larger values of the density $h$ of
empty sites, a regime where the cultural dynamics shows strikingly
different features. At very low values of the initial cultural
diversity $q$ (so that cultural convergence is strongly favored),
the asymptotic states are characterized by low values of the order
parameter $\langle S_{\text {max}}\rangle /N$. The reason for the
absence of cultural globalization in this regime is the formation
of disconnected monocultural domains, a fact that requires values
of the density $1-h$ of cultural agents (at least) close to (or
below) the site percolation threshold value for the square
lattice (0.593). This new kind of macroscopic multicultural state is thus
of a very different nature from the multicultural phase of the
original Axelrod model ($h=0$). The values of the order parameter
in this {\em fragmented} phase, represented in Fig.\ \ref{fig2}a as a function of $q/N$ with $h=0.5$ and $T=0.7$ and for several values of $L$, decrease with increasing lattice
size, and the expectation is that the order parameter vanishes in
the thermodynamical limit, because the largest cluster size below
percolation should be independent of the lattice size.

The increase of $q$ from the very small values that correspond to
the fragmented multicultural phase has the seemingly paradoxical
effect of increasing the order parameter $\langle S_{\text
{max}}\rangle /N$ values, {\em i.e.}, the increase of the initial
cultural disorder promotes cultural globalization. To understand
this peculiar behavior, one must consider the effect of the
increase of $q$ in the initial mobility of the agents. One expects
that the higher the value of $q$ is, the lower the initial values
of the cultural overlap $\omega_{ij}$ among agents are, and then
the higher the initial mobility of agents should be. Under
conditions of high mobility, the processes of local cultural
convergence are slower than the typical time scales for mobility,
so that the agents can easily move before full local consensus can
be achieved, propagating their common features, and enhancing the
social influence among different clusters. In other words, the
attainment of different local consensus in disconnected domains is much
less likely to occur, and one should expect the coarsening of a
dominant culture domain that reaches a higher size. A
straightforward prediction of this argument is that one should
observe higher values of $\langle S_{\text {max}}\rangle /N$ for
higher values of the intolerance $T$, because agents mobility is
an increasing function of this parameter (see eq.
(\ref{mobility})). The numerical results shown in Fig.\ \ref{fig2}b for
different values of $T$ and $h=0.5$ nicely confirm this
prediction, in support of the consistency of the previous
argument. Interestingly, for high values of the intolerance $T$,
an almost full degree of cultural globalization is reached, as
indicated by the values $\langle S_{\text {max}}\rangle /N \simeq
1$ of the order parameter. In those final states almost all agents
belong to a single connected monocultural cluster. On the contrary, for very low values of $T$ when mobility is not enhanced, multiculturalism prevails for the whole range of $q$ values.

To characterize the passage from the multicultural fragmented
phase to global consensus with increasing initial cultural
diversity, we have computed the histograms of the values of
$S_{\text {max}}/N$ at values of $q$
where the order parameter increases. The histograms display the
bimodal characteristics of a first-order transition. In a fraction
of realizations, the transient mobility is able to spread social
influence among the clusters so that global consensus is finally
reached. This fraction increases with $q$,
to the expense of the fraction of realizations where fragmented
multiculturality is reached. 

\begin{figure}
\begin{center}
\epsfig{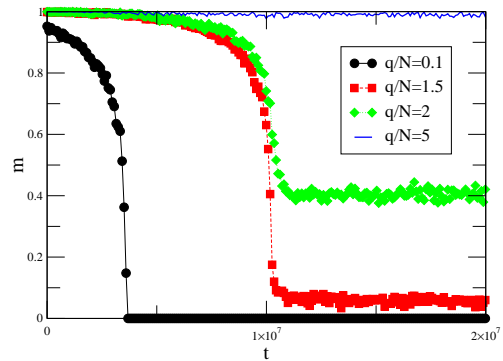}
\end{center}
\caption{(color online) Average mobility $m$ versus time $t$ for $h=0.5$, $L=30$, $T=0.7$ and different values of
the scaled initial culture diversity $q/N$ as indicated. Unlike the
other figures, in this case each curve represents the results of a
single realization. See the text for further details.}
\label{fig3}
\end{figure}
%\end{center}

Further increase of the initial cultural diversity $q$ enhances
the likelihood of agents sharing no cultural feature with anybody
else in the finite population. The presence of these culturally
"alien" agents decreases the value of the order parameter and the
increase of their number with $q$ is concomitant with the
transition to multiculturality in the original Axelrod model (as
well as here, for finite populations). We see in Fig.\ \ref{fig2}b that the
increase of the intolerance parameter $T$ shifts this transition
to higher values of $q/N$, in agreement with the enhancement of
the convergence to globalization that $T$ produces via mobility,
as discussed above. Each alien agent has, at all times, a mobility $m_i=1$, and the
average mobility cannot decrease in time to zero value when they
appear. In other words, the asymptotic states of the cultural
dynamics are no longer characterized by $m=0$. The time evolution of the average mobility $m$ for particular realizations at $h=0.5$, $T=0.7$, $L=30$ and different values of $q/N$ is shown in Fig\ \ref{fig3}. The value of $q/N$ beyond which the stationary average mobility is larger than zero signals the appearance of these alien cultural agents.

%\begin{center}
\begin{figure}[t]
\begin{center}
\epsfig{file=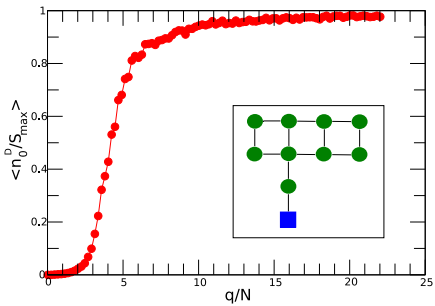, width=6.5cm, angle=0}
\end{center}
\caption{(color online) As a quantitative measure of this erosion phenomenon we plot here the
stationary value of the averaged fraction $n^D_0/S_{max}$ of isolated 
individuals of the dominant culture versus $q/N$, for $h=0.5$, $T=0.7$, 
and $L=30$. The inset shows an illustrative configuration where erosion can take place.}
\label{fig4}
\end{figure}
%\end{center}

In addition, the restless character of the alien agents has an important effect on
the geometry of the dominant culture, namely its {\em erosion}. As
an illustrative example, let us consider the situation represented
in the inset of Fig.\ \ref{fig4}, in which an agent $i$ of the dominant culture is placed
at the frontier of a cluster, having a single neighbor of his
kind, and assume that an alien agent $j$ has moved recently to one
of the empty neighboring sites of $i$. When agent $i$ is chosen
for an elementary dynamical step, there is a probability $1/2$ of
choosing agent $j$ for an imitation trial. As $\omega_{ij}=0$, and
then $\bar{\omega}_i=1/2$, the agent $i$ will move from there to a
randomly chosen empty site whenever the intolerance parameter is
$T>1/2$. We see that, for this particular situation, the erosion
of the dominant culture cluster will occur with probability one
half.

Note that the erosion of the dominant culture cluster does not
change the size $S_{\text {max}}$ of the dominant culture. It
simply breaks it up into separate domains, some of them consisting
of single (isolated) individuals. These isolated members of the
dominant culture will eventually adhere to domains, to be at a
later time again exposed to erosion, and so on. Therefore the
shape and number of domains of the dominant culture (as well as
that of the other ones), fluctuate forever. The number $n^D_0$ of
isolated dominant culture agents reaches a stationary value that results from the balance between erosive and adhesive processes. To quantify the strength of the eroding activity of cultural
minorities we show in Fig.\ \ref{fig4} the stationary value of the averaged
fraction $\langle \frac{n^D_0}{S_{max}} \rangle$ of isolated individuals of
the dominant culture versus the scaled initial cultural diversity, for
$h=0.5$, $T=0.7$, and $L=30$. Soon after the transition from the
fragmented multicultural phase to globalization occurs, erosion
increases dramatically, largely contributing to the large values of
the stationary mobility $m$ that characterize the multicultural states
in the model here introduced.
%Axelrod-Schelling model.}

%\section{Conclusions}

In summary, the introduction of agents mobility through this segregation mechanism into the Axelrod cultural
dynamics leads to an enhancement of the convergence to cultural
globalization for small densities of empty sites, while for larger
densities a new type of multicultural fragmented phase appears at
low values of the initial cultural diversity $q$, followed by a
new transition to globalization for increasing values of $q$ that
is triggered by the increase in the initial mobility. Moreover, in the
genuine Axelrod transition from global consensus to polarization,
the shape and number of cultural domains are here dynamically
fluctuating by the competitive balance of erosive and adhesive
processes associated to the agents mobility.

\begin{acknowledgments}

Y. M. is supported by MICINN (Spain) through the Ram\'on y Cajal
Program.  This work has been partially supported by MICINN through
Grants FIS2006-12781-C02-01, and FIS2008-01240, and by
Comunidad de Arag\'on (Spain) through a grant to FENOL group.

\end{acknowledgments}

\end{document}